\documentclass[epj,english,twocolumn]{webofc}
\usepackage[varg]{txfonts}   
\usepackage{hyperref}        

\newcommand{\al}{\alpha}

\newcommand{\az}{\varphi}
\newcommand{\ro}{\rho}

\newcommand{\oeq}{\begin{equation}}
\newcommand{\ceq}{\end{equation}}
\newcommand{\oeqn}{\begin{eqnarray}}
\newcommand{\ceqn}{\end{eqnarray}}

\renewcommand{\>}{\rangle}
\newcommand{\<}{\langle}
\renewcommand{\(}{\left(}
\renewcommand{\)}{\right)}

\newcommand{\stf}{\,\,\,}
\newcommand{\sdf}{\,\,}
\newcommand{\stb}{\!\!\!}

\newcommand{\kfi}{|\phi \>}

\newcommand{\bfi}{\<\phi |}

\newcommand{\oQ}{\hat{Q}}

\newcommand{\oH}{\hat{H}}

\newcommand{\of}{\hat{f}}

\newcommand{\oF}{\hat{F}}

\newcommand{\hb}{\hbar}

\newcommand{\vr}{{\bf r}}

\newcommand{\vp}{{\bf p}}

\woctitle{Heavy Ion Accelerator Symposium 2013}
\begin{document}
\title{Probing quantum many-body dynamics in nuclear systems}
%
%

\author{C. Simenel\inst{1}\fnsep\thanks{\email{cedric.simenel@anu.edu.au}} \and
        M. Dasgupta\inst{1} \and
        D. J. Hinde\inst{1} \and
        A. Kheifets\inst{2} \and
        A. Wakhle\inst{1}}

\institute{Department of Nuclear Physics, Research School of Physics and Engineering,  Australian National University, Canberra, Australian Capital Territory 0200, Australia
\and
           Atomic and Molecular Physics Laboratory, Research School of Physics and Engineering,  Australian National University, Canberra, Australian Capital Territory 0200, Australia
                     }

\abstract{
Quantum many-body nuclear dynamics is treated at the mean-field level with the time-dependent Hartree-Fock (TDHF) theory. 
Low-lying and high-lying nuclear vibrations are studied using the linear response theory. 
The fusion mechanism is also described for light and heavy systems. 
The latter exhibit fusion hindrance due to quasi-fission. 
Typical characteristics of quasi-fission, such as contact time and partial symmetrisation of the fragments mass in the exit channel, are reproduced by TDHF calculations.
The (multi-)nucleon transfer at sub-barrier energies is also discussed.
}
\maketitle
\section{Introduction}
\label{intro}
The quantum many-body problem is common to all fields aiming at describing complex quantum systems of interacting particles. Examples range from quarks and gluons in a nucleon to macromolecules such as fullerenes. 
Breakthroughs in one field may have a strong impact in others. 
For instance, the development of the BCS theory to describe superconducticity \cite{bar57} has been crucial to understand some properties of atomic nuclei due to pairing correlations. 
Another example is the description of low-energy fusion with multi-channel tunnelling \cite{das98} which is now used to investigate dissociative adsorption of molecules in surface science \cite{hag12}.

Nuclear systems are interesting examples of many-body systems where up to about 500 nucleons (in the case of actinide collisions) may interact. What makes the dynamics of nuclear systems special to test quantum many-body theories is their  almost complete isolation from external environments. 
Indeed, the coupling of a system to its environment induces a decoherence process which is responsible for the quantum to classical transition \cite{zur91,joo03}. 
Table \ref{tab1} gives some properties of usual microscopic many-body systems. 
Depending on their size and the native time scale associated to their dynamics, these systems may interact with electromagnetic (EM) fields and with the particles of the surrounding gas. 
The outcome of the collision between atomic nuclei, however, is usually determined before any decoherence process takes place. 
This is due to their small size (few fm) and the fact that the  interaction times are of the order of few zeptoseconds, i.e., much shorter than typical times for gamma emission. 
As a result, the coherent superposition of quantum states which is built up during the collision is preserved during times which are typically much longer than the collision time itself.
Heavy-ion collisions are then ideal to investigate fundamental aspects of quantum physics, such as collective motion \cite{boh75}, tunnelling and dissipation \cite{cal81}, coupled channels \cite{hag12}, and entanglement \cite{lam76}.\par

\begin{table}[!t]
\centering
\caption{Properties of typical many-body systems and the environment they may interact with, inducing possible quantum to classical transition. As a result, the equations of motion used to describe the dynamics of the system is derived from a quantum or classical mechanics approach.}
\label{tab1}       
\begin{tabular}{l|lll}
\hline
& Nuclei & Atoms & Molecules  \\\hline
Size & $\sim10^{-14}$m & $\sim10^{-10}$m& $\sim10^{-9}$m \\
Time scale & $10^{-21}$s=1zs& $10^{-18}$s=1as & $10^{-15}$s=1fs \\
Environment & none& EM& EM+gas\\\hline
Eq. of& \textit{Quantum} & \textit{Quantum} & \textit{Classical} \\
motion& & \textit{(Classical)} & \textit{(Quantum)} \\\hline
\end{tabular}
\end{table}

Describing nuclear dynamics and predicting the outcome of heavy-ion collisions is very challenging as several  mechanisms may occur. Ideally, the same theoretical model should be able to describe vibrational and rotational motions and all the reaction outcomes, e.g., (in)elastic scattering, multi-particle transfer, and fusion. A good starting point is to consider that the particles evolve independently in the mean-field generated by the ensemble of particles. This leads to the well known time-dependent Hartree-Fock (TDHF) theory proposed by Dirac \cite{dir30} which has been applied to many nuclear systems in the past decade (see Ref. \cite{sim12} for a review).\par

In this contribution, we present recent applications of the TDHF approach to nuclear dynamics, from vibrations to heavy-ion collisions around the Coulomb barrier.
The TDHF approach is first presented. 
Low-lying collective vibrations and giant resonances are then discussed as a first application. 
This is followed by a study of the fusion mechanism in light systems.
We then discuss the quasi-fission mechanism responsible for fusion hindrance in heavy systems. 
Finally, we investigate nucleon transfer at sub-barrier energies.

\section{The TDHF approach}

The first  TDHF numerical calculations in nuclear physics were performed in the mid 70's  \cite{bon76,neg82}.
Standard applications of the TDHF approach in nuclear physics are performed with  an effective interaction derived from an energy density functional (EDF).
The EDF is the only phenomenological ingredient of the model.
In practice, Skyrme EDFs are used~\cite{sky56}. 
They are usually adjusted to fit only a few nuclear properties, e.g., infinite nuclear matter, radii and masses of few doubly magic nuclei~\cite{cha98}. 
The many-body wave function is chosen to be an antisymmetrised independent particle state at any time~$t$. 
The Pauli principle is then exactly treated during time evolution. 
It is crucial to treat properly  Pauli blocking, which prevents nucleon-nucleon collisions and makes the mean-field approximation valid at low energies.

Modern TDHF calculations are fully consistent in the sense that they treat static properties {\it and} dynamics on the same footing. 
A proper treatment of nuclear structure is indeed crucial for near-barrier reactions which are highly sensitive to couplings between the relative motion and internal degrees of freedom.

\subsection{The TDHF formalism}

The TDHF equation is derived from the action  (see, e.g., appendix A of Ref. \cite{sim12})
\oeq
S \equiv S_{t_0,t_1}[\phi] = \int_{t_0}^{t_1} \stb d t \stf \<\phi (t)| \( i\hb \frac{d }{{d} t} - \oH \) |\phi(t)\>
\ceq
where $\oH$ is the Hamiltonian of the system and the state of the $A$ independent particles is constrained to be a Slater determinant $\kfi$ at any time. 
In the framework of the energy density functional theory, the expectation value of the Hamiltonian is written as a functional of the one-body density matrix $\ro$: $E[\ro] = \bfi \oH \kfi$. 
For independent particle systems, $\ro$ contains the same information as $|\phi\>$.
It is expressed as a function of the occupied single-particle wave-functions $\varphi_i$ as
\begin{equation}
\rho(x,y) = \sum_{i=1}^{A} \sdf  \varphi_i(x)\sdf \varphi_i^*(y),
\end{equation}
where $x\equiv (\vr s q)$ describes all the single-particle degrees of freedom (position $\vr$, spin $s$ and isospin $q$). 

Solving the variational principle  $\delta S = 0$, we get  a set of Schr\"odinger-like equations for each single-particle wave-function
\oeq
  i\hb \sdf \frac{ d }{ d t} \az_i(x,t) = \int \stb  d y \stf h[\ro(t)](x,y) \sdf \az_i(y,t), \label{eq:TDHF_az}
 \ceq
where the Hartree-Fock single-particle Hamiltonian $h[\rho]$ reads
\begin{equation}
h[\rho](x,y) = \frac{\delta E[\rho]}{\delta \rho(y, x)}.
\end{equation}
The system of equations (\ref{eq:TDHF_az}) are coupled due to the self-consistency of the HF Hamiltonian. 
The TDHF equation can be expressed from these equations in a compact form as
\begin{equation}
i\hbar \frac{\partial}{\partial t} \rho = \left[h[\rho],\rho\right].
\label{eq:tdhf}
\end{equation}

\subsection{Numerical details}

Three-dimensional codes have recently been developed to solve the TDHF equation for the dynamics of realistic nuclear systems \cite{kim97,mar05,nak05,uma05,seb09}.
Here, the TDHF solution is obtained from the set of equations (\ref{eq:TDHF_az}) which are solved iteratively in time on a Cartesian grid with hard-boundary conditions using  the  {\textsc{tdhf3d}} code \cite{kim97}.
The SLy4 \cite{cha98} parametrisation of the Skyrme EDF is used in calculations of vibrational modes of the nucleus.
To describe heavy-ion collisions, we use the SLy4$d$ parametrisations \cite{kim97} in which center of mass corrections are not accounted for in the fitting procedure.
The lattice spacing is $\Delta x=0.8$~fm and the time step is $\Delta t=1.5\times10^{-24}$~s.

\subsubsection{Dynamics of one nucleus}

The study of nuclear vibrations is a common application of the TDHF theory (see Refs. \cite{mar05,uma05,nak05,sim03,rei07,sim09,fra12} for recent studies with three-dimensional codes). 
The ground-state of the nucleus is first obtained from a static HF code with the same EDF as in the dynamical calculation. 
A time-dependent perturbation $V(t)$ is then applied to the nucleus. 
Usually, this perturbation is proportional to a delta-function $\delta(t)$ and can then be applied using a boost operator on the ground-state at the initial time:
\oeq
|\phi(0)\>=\exp(-i\varepsilon \oF)\sdf|\phi^{HF}\>,
\ceq
where $\oF=\sum_{i=1}^A\of_{i}$ is a one-body operator and $\varepsilon$ is the boost velocity. 
In practice, the boost is applied on the single-particle wave-functions according to
\oeq
|\az_i(0)\>=\exp(-i\varepsilon \of)\sdf|\az^{HF}_i\>.
\ceq

In a second step, the time response of the wave-functions is computed over a finite time of typically $10-20$~zs.
The case of unbound states should be considered with care as emitted particles bounce back from the edge of the wall due to the hard boundary conditions.
The reflected particles can then induce a spurious motion when interacting with the system \cite{rei06}.
Absorbing boundary conditions should then be considered to absorb the emitted particles at the edge of the box \cite{nak05,rei06,par13}.
Note that one may be interested in the information contained in these emitted particles, e.g., to assess the microscopic composition of a giant resonance \cite{ave13}. 
In this case, however, large boxes need to be considered which prevent the use of three-dimensional codes. 

\subsubsection{Collision of two nuclei}

The HF ground state of the nuclei are first computed with a static HF code. 
Then, they are placed in a large Cartesian box at a distance $D_0$ which has to be large enough (typically $D_0\sim30-50$~fm)
to account for Coulomb excitation in the entrance channel.

The initial momenta of the nuclei $\vp_{\al=1,2}$ are determined assuming a Rutherford trajectory prior to the initial time. 
A Galilean boost of the form 
\oeq
\az_i(\vr sq;t=0)=\exp(\frac{i}{\hb}\vp_\al\cdot\vr)\az^{HF}_i(\vr sq)
\ceq
is applied at the initial time to the single-particle wave-functions of the nucleus $\al$.

The evolution is then performed over a finite time (usually several zs) during which some observables (e.g., distance between the nuclei, multipole moments...) are computed. 
The outcome of the collision (e.g., fusion or  transfer products) can then be analysed using the wave function at the final time. 

\subsection{Beyond TDHF approaches}

The TDHF approach gives only classical trajectories for the time-evolution  and expectation values of one-body observables. 
In particular, it does not include tunnelling of the many-body wave function.
One way to overcome this difficulty is to determine the nucleus-nucleus potential from TDHF trajectories using the Density-Constrained TDHF  \cite{uma06a} or the Dissipative-Dynamics TDHF method \cite{was08}.
The sub-barrier fusion probability can then be determined by integrating the Schr\"odinger equation with this potential. 

TDHF calculations may also underestimate fluctuations of one-body observables~\cite{das79,bal81}.
Such fluctuations have been computed recently at the time-dependent random phase approximation (TDRPA) limit \cite{bro08,sim11} using a prescription proposed by Balian and V\'en\'eroni~\cite{bal84}, or using stochastic extensions \cite{was09}.
In particular, it has been shown that the width of the mass and charge distributions of fragments produced in deep-inelastic collisions are strongly underestimated at the TDHF level, while TDRPA predictions are in reasonable agreement with experimental data \cite{sim11}. 

Pairing correlations responsible for superfluidity in nuclei have been included to study  vibrations in nuclei~\cite{ave08,eba10,ste11,asc12,sca12} and more recently pair transfer reactions \cite{sca13}. 
In particular, it has been shown that these correlations are responsible for pair vibrations \cite{ave08} and for an enhancement of pair transfer cross-sections \cite{sca13}. 

Another type of correlations are induced by the collision term which is expected to affect nuclear dynamics at energies well above the Coulomb barrier, where Pauli blocking becomes less efficient in preventing  nucleon-nucleon collisions. 
Recent calculations based on the time-dependent density-matrix formalism \cite{wan85} have been performed to investigate the role of the  collision term on nuclear vibrations \cite{toh01} and collisions \cite{toh02,ass09}.

Although the increase of computational power has enabled realistic applications of some of these beyond-TDHF approaches, the numerical cost remains usually too high for systematic studies. 
The TDHF theory then remains a popular approach, in particular thanks to its successes in the past decade. 
In the following, we present some recent applications at the TDHF level.

\section{Applications of the TDHF approach to nuclear dynamics}

\subsection{Collective vibrations \label{sec:vib}}

Atomic nuclei exhibit a large variety of  vibrations, from low-lying collective modes to giant resonances (GR), which can be modeled by the TDHF theory. 
In fact, basic properties of vibrations such as energy and strength can be computed using the linear response theory, which, applied within the TDHF framework, is equivalent to the Random Phase Approximation (RPA).

An interesting aspect of using TDHF codes, however, is that one is not limited to small amplitude vibrations, unlike RPA, allowing for investigations of non linear effects in collective motions.
In particular, anharmonicities and coupling between collective modes have been investigated~\cite{sim03,rei07,sim09}. 

In the present work, however, we restrict the discussion of nuclear vibrations to the linear response. 
After a brief presentation of the linear response theory, we present applications to low-lying collective modes and to giant resonances. 

\subsubsection{Linear response theory}

In the linear response theory, the evolution of the wave-function after a boost of the form
\oeq
|\Psi(t=0)\> = e^{-i\varepsilon \oQ } |\Psi_0\>,
\ceq
where $|\Psi_0\>$ is the ground state,
is used to determine the evolution of the operator $\oQ$ which, in the first order in the boost velocity $\varepsilon$, reads
\oeqn
\Delta Q(t)&=&\<\Psi(t)|\oQ|\Psi(t)\>-\<\Psi_0|\oQ|\Psi_0\> \nonumber \\
&\simeq& -2\varepsilon\sum_\nu |q_\nu|^2\sin \omega_\nu t,
\label{eq:linresp}
\ceqn
where  $q_\nu=  \<\Psi_\nu| \oQ |\Psi_0\>$ is the transition amplitude between 
the ground state and the eigenstate $|\Psi_\nu\>$ of the Hamiltonian with eigenenergy $E_\nu=E_0+\hb \omega_\nu$.

The strength function is then defined as
\begin{eqnarray}
R_{Q}(\omega) &=&\lim_{\varepsilon\rightarrow 0} \frac{-1 }{\pi \varepsilon}\,
\int_{0}^{\infty}  dt\, \Delta{Q}(t) \, \sin (\omega t). \label{eq:strengthlin} \\
 &=& \sum_\nu \, |q_\nu |^2 
 \delta (\omega - \omega_\nu). \label{eq:strengthfinal}
\end{eqnarray}
Note that the strength function is usually represented as a function $S_Q(E=\hb\omega)$ according to
\begin{equation}
S_Q(E) = R_Q(\omega)/\hb = \sum_\nu \, |q_\nu |^2 
 \delta (\hb\omega - \hb\omega_\nu).
 \end{equation}

\subsubsection{Low-lying collective modes \label{sec:lowvib}}

\begin{figure}[!tb]
\centering
\includegraphics[width=7cm]{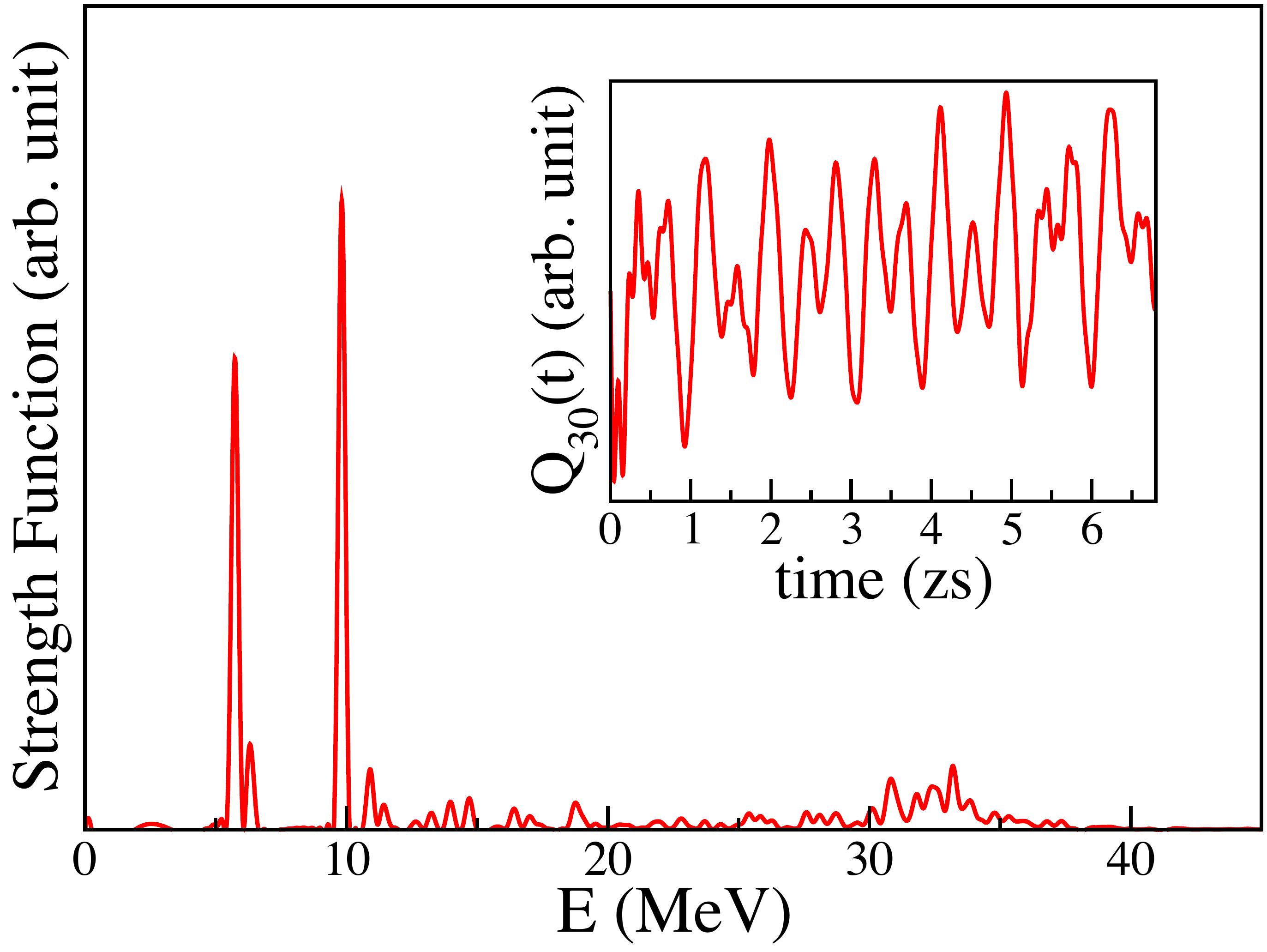}
\caption{Strength function of the octupole response in $^{48}$Ca.
The associated TDHF time-evolution of the octupole moment is shown in the inset. }
\label{fig:48Ca3-}   
\end{figure}

Figure \ref{fig:48Ca3-} shows the strength function associated with the octupole moment 
\begin{equation}
Q_{30} = \sqrt{\frac{7}{16\pi}}\int d^3r \rho(\mathbf{r}) \left[2{x}^3-3{x}\left({y}^2+{z}^2\right) \right]
\label{eq:Q3}
\end{equation}
in $^{48}$Ca, where $\rho(\vr)=\sum_{s,q} \rho(\vr s q, \vr s q)$ is the local part of the one-body density matrix.   
The time-evolution of the octupole moment  following the octupole boost is shown in the inset. 

Two collective modes at 5.7 MeV and 9.8 MeV are seen, exhausting $8\%$ and $18\%$ of the energy weighted sum rule (EWSR), respectively. 
The energy of the first peak overestimates the experimental value of the energy of the $3^-_1$ state, $E_{3^-_1}=4.507$~MeV, by about $27\%$. 
A similar factor ($30\%$) was obtained in the $^{208}$Pb nucleus with similar calculations \cite{sim12}. 

The overestimation of the energy of the $3^-_1$ collective state might be a general feature of the SLy4 parametrisation. 
Systematic calculations of low-lying octupole modes across the nuclear chart and using various Skyrme interactions should then be performed. 

\subsubsection{Giant resonances}

Some strength can be observed in Fig.~\ref{fig:48Ca3-} above 30~MeV, which could be associated with a high-energy octupole resonance (HEOR).
HEOR are a type of giant resonance (GR) for the octupole motion, with typical energies of $110/A^{1/3}$ \cite{car80}. 
Other types of giant resonances have been observed experimentally, such as the isovector giant dipole resonance (GDR), the isoscalar giant monopole resonance (GMR, also called the breathing mode) and the isoscalar giant quadrupole resonance. 

Giant resonances are usually unbound and decay by particle emission, leading to an escape width in the GR spectra. 
Other contributions to the width of GR are the Landau damping \cite{pin66} and the spreading width. 
Only the escape width and the Landau damping are included in the TDHF approach. 
As a result, TDHF calculations of giant resonances usually underestimate the total width of the GR. 

\begin{figure}[!tb]
\centering
\includegraphics[width=6cm]{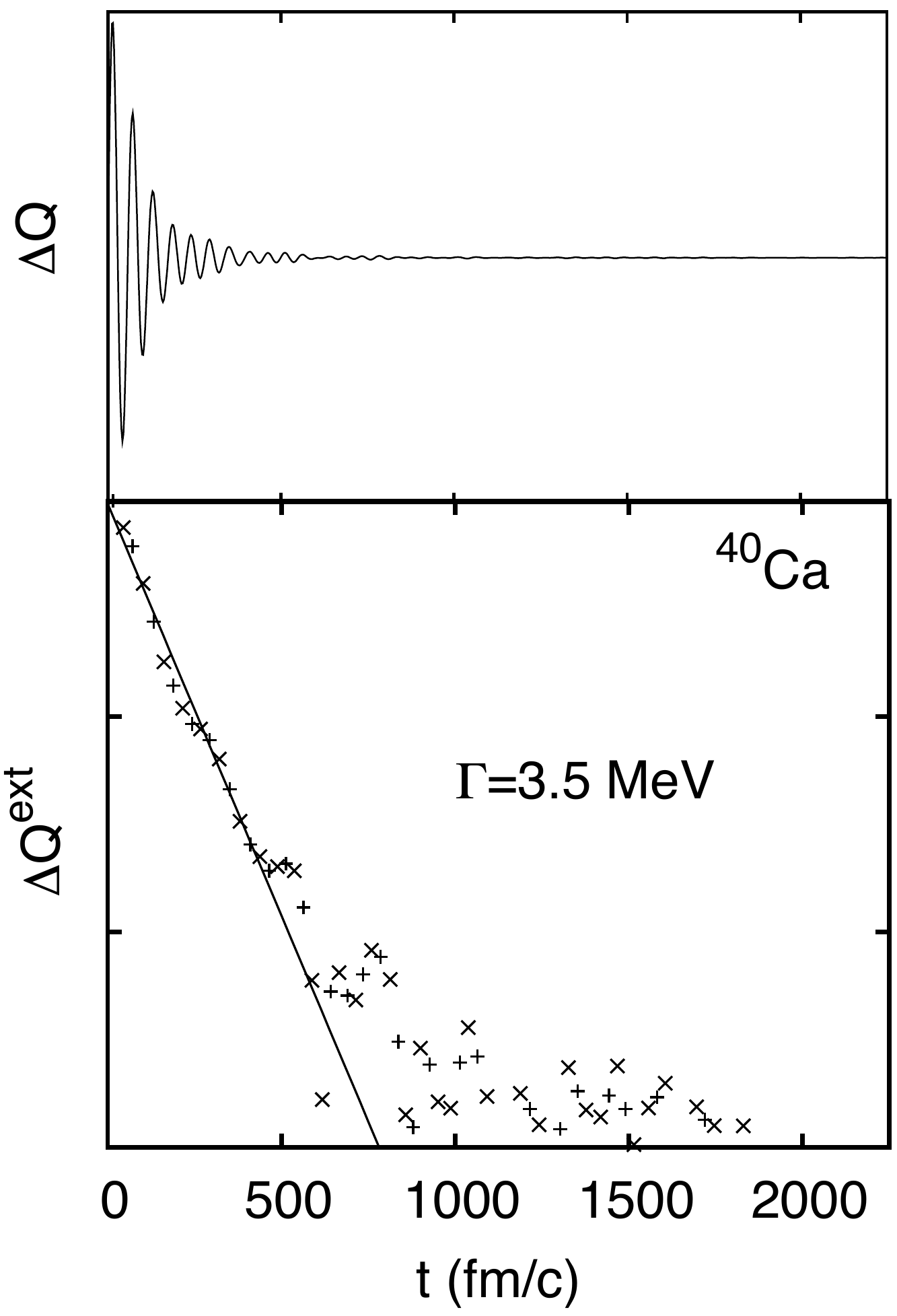}
\caption{(upper panel) Time evolution of the monopole moment in $^{40}$Ca after a monopole boost in the linear regime.
(lower panel) Time evolution of the amplitude of the monople oscillations (''$\times$''=maxima, ''$+$''=minima).}
\label{fig:40CaQ0}   
\end{figure}

An example of a TDHF calculation of the monopole response to a monopole boost in $^{40}$Ca is shown in Fig.~\ref{fig:40CaQ0}. 
The monopole moment is defined as 
\begin{equation}
 Q_0 = \sqrt{\frac{1}{4\pi}}\int d^3r \rho(\mathbf{r}) {r}^2 .
\label{eq:Q0}
\end{equation}
To avoid spurious contributions from particles reflected on the hard box boundary, the calculations are performed on a large spherical grid of 600~fm with the   \textsc{tdhfbrad} code (without pairing) of Ref.~\cite{ave08,ave13}. 

It can be seen from the top of Fig.~\ref{fig:40CaQ0} that the monopole oscillations are damped.
This damping is due to the direct decay of the GMR by nucleon emission. 
Fitting the extrema $\Delta Q^{ext}(t)$ of the monopole moment with $\exp(\frac{-\Gamma t}{2\hb})$ gives a width $\Gamma\simeq3.5$~MeV (see lower panel of Fig.~\ref{fig:40CaQ0}).

\begin{figure}[!tb]
\centering
\includegraphics[width=6cm]{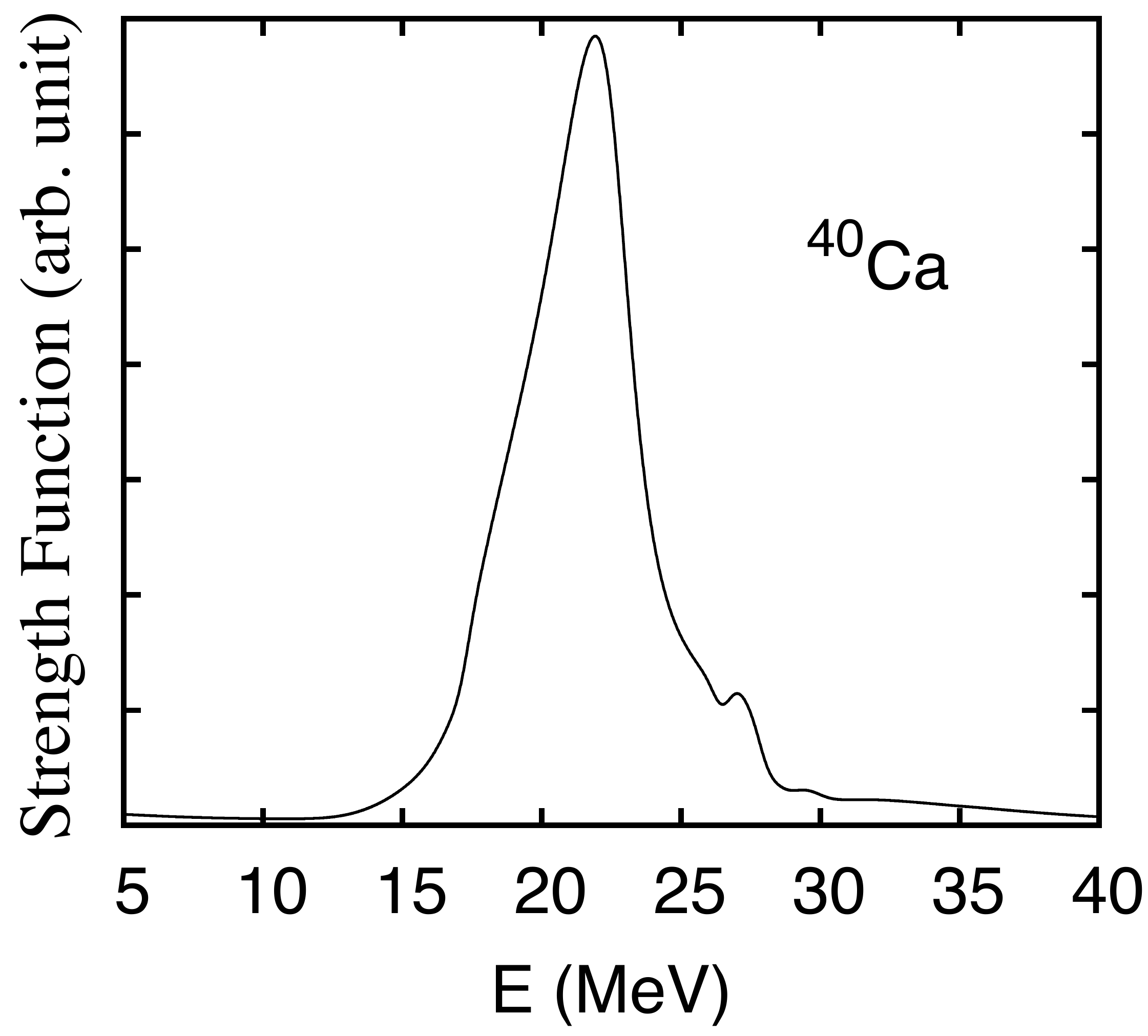}
\caption{Strength function of the GMR in $^{40}$Ca.}
\label{fig:40CaS0}   
\end{figure}

The strength function of the GMR has been computed from the time response of the monopole moment shown in the upper panel of Fig. \ref{fig:40CaQ0}. 
The result is shown in Fig.~\ref{fig:40CaS0}. 
The calculated energy of the GMR $E_{GMR}^{TDHF}\simeq 22.1$~MeV slightly overestimates the experimental value $E_{GMR}^{exp}=19.2\pm0.4$ \cite{you01}. 

The calculated width of the GMR extracted from the FWHM of the strength in Fig.~\ref{fig:40CaS0} is $\Gamma_{tot}^{TDHF}4.5$~MeV.
This value is 1~MeV larger than the one calculated from Fig.~\ref{fig:40CaQ0} because of the Landau damping. 
Overall, this width is in good agreement with the experimental value $\Gamma_{tot}^{exp}\simeq4.9\pm0.6$ \cite{you01}.
This indicates that the spreading width, which is neglected in TDHF, only slightly affects the total width of the GMR in $^{40}$Ca. 
This is usually the case for the GMR in light nuclei (see, e.g., Ref.~\cite{ave13} for a study of the GMR in $^{16}$O). 
However, the spreading width becomes dominant in the GMR of heavy nuclei (see, e.g., a study of the GMR in $^{208}$Pb in Ref.~\cite{sim12}). 

Finally, it is worth mentioning that recently similar TDHF calculations have been performed to study the direct decay from the GMR in $^{16}$O \cite{ave13}. 
The spectrum of emitted nucleons exhibits structures which reflect the single-particle structure of the nucleus. 
To some level, GR direct decay by nucleon emission could then be used to investigate the shell levels in light nuclei. 

\subsection{Fusion mechanism}

The merging of collision partners into a compound system is a complex, highly non-linear, and irreversible process. 
It is strongly coupled to internal structures of the colliding partners resulting from their quantum nature, as well as other reaction mechanisms such as (multi)-nucleon transfer. 

\subsubsection{Path to fusion in light systems}

\begin{figure}[!tb]
\centering
\includegraphics[width=7cm]{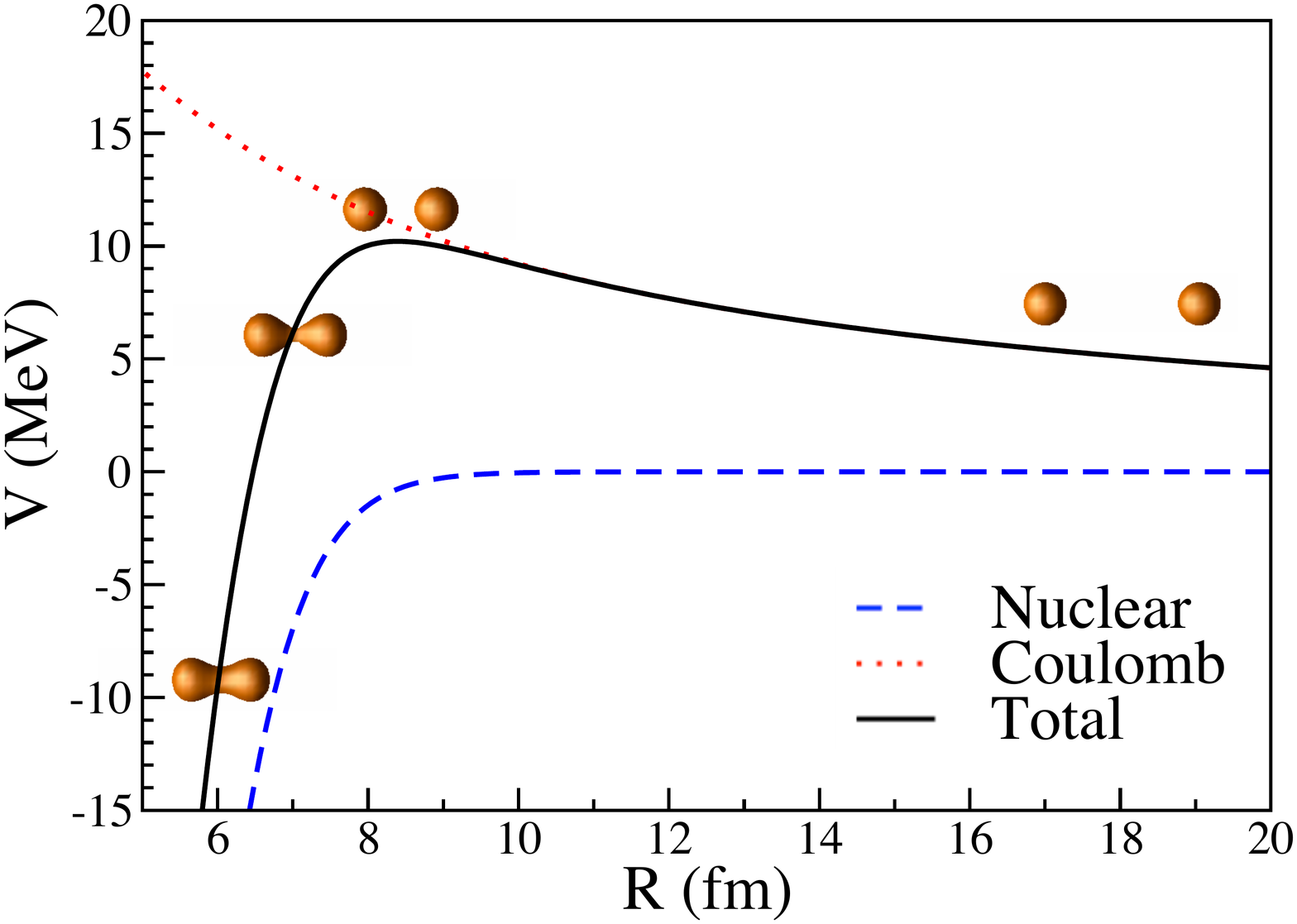}
\caption{Nucleus-nucleus potential as a function of the distance between the centers of mass of the fragments in the $^{16}$O+$^{16}$O system from the frozen-HF technique. 
The snapshots show densities at half the saturation density at $R=18$, 8.4, 7 and 6~fm from a TDHF calculation at $E_{c.m.}=12$~MeV with the SLy4d interaction. } 
\label{fig:Frozen_O+O}   
\end{figure}

The reaction mechanisms depend on the characteristics of the nuclei and in particular on their mass and charge. 
Bringing light nuclei into contact is usually sufficient to cause them to fuse. 
This is illustrated in Fig.~\ref{fig:Frozen_O+O} where the nucleus-nucleus potential in $^{16}$O+$^{16}$O is plotted as a function of the relative distance $R$ between the nuclei. 
We define $R$ as the distance between the centers of mass of the matter distribution on each side of the neck. 
The potential is obtained from the frozen-HF technique, where the energy of the system is computed from the EDF considering HF densities at a fixed distance. 
As a result, the fusion barrier, generated from the competition between the nuclear and the Coulomb potentials, is reached at $R\simeq8.4$~fm. 

The $^{16}$O+$^{16}$O system has been recently investigated with modern TDHF codes \cite{sim13}.
Fig.~\ref{fig:Frozen_O+O} also shows snapshots of the density at different distances from a TDHF calculation at about $E_{c.m.}=12$~MeV.
We see  that the two nuclei are still well separated when the fusion barrier is reached. 
A neck is formed inside the barrier, at $\sim7$~fm. 
At $R\sim6$~fm, the size of the neck increases and the fragments merge. 

It is well known that fusion around the barrier is highly sensitive to the structure of the colliding nuclei, in particular to their low-lying vibrational modes (like those described in section \ref{sec:lowvib}) and to their rotational states \cite{das98}.
The standard approach to describe the coupling between the relative motion and these internal excitations is the coupled-channel formalism \cite{hag12}. 
However, TDHF calculations have also been used to describe the effect of rotation and deformation on fusion \cite{sim04,uma06b}. 
One advantage of the TDHF approach is that the coupling between the internal structure (e.g., low-lying vibrational and rotational modes) and the relative motion is included at all orders at the mean-field level. 
In particular, the energy of the states and their transition amplitudes are not input parameters of the calculations.

In addition, collective vibrations can appear built on any shape of the system during its path to fusion. 
In particular, the pre-equilibrium giant dipole resonance, which is excited in $N/Z$ asymmetric collisions, has been studied in detail with modern  TDHF codes \cite{sim01,sim07,obe12}. 
It has been shown that the properties of the pre-equilibrium GDR could be used to infer the characteristics of the system on its path to fusion. 
For instance, a lowering of the pre-equilibrium GDR energy in comparison with the GDR in a spherical nucleus could be related to a large deformation of the compound system \cite{sim07}.

\subsubsection{Fusion hindrance and quasi-fission in heavy systems}

As mentioned earlier, the path to fusion strongly depends on the mass and charge of the nuclei. 
In fact, unlike light systems, for the fusion of heavy systems contact between the reactants is clearly not sufficient. 
Indeed, the latter exhibit fusion hindrance due to the quasi-fission mechanism. 
Mass flow between the reactants occurs, leading to re-separation of more symmetric fragments in the exit channel \cite{boc82,tok85}.
A good understanding of the competition between the fusion and quasi-fission mechanisms is expected to be of great help to optimise the formation and study of heavy and superheavy nuclei. 

The quasi-fission process is associated with typical contact times between the fragments up to $\sim20$~zs \cite{boc82,tok85,rie11}. 
Although these times are longer than the time needed to overcome the barrier in light systems (usually less than 2~zs, see, e.g., Ref.~\cite{leb12}), they are several orders of magnitude shorter than the statistical fission time of the heavy compound nuclei \cite{mor08,fre12}. 
In fact, quasi-fission occurs before the equilibration of the degrees of freedom (shape, charge and mass asymmetry, angular distribution...) and the quasi-fission products keep a memory of the entrance channel. 

\begin{figure}[!tb]
\centering
\includegraphics[width=7cm]{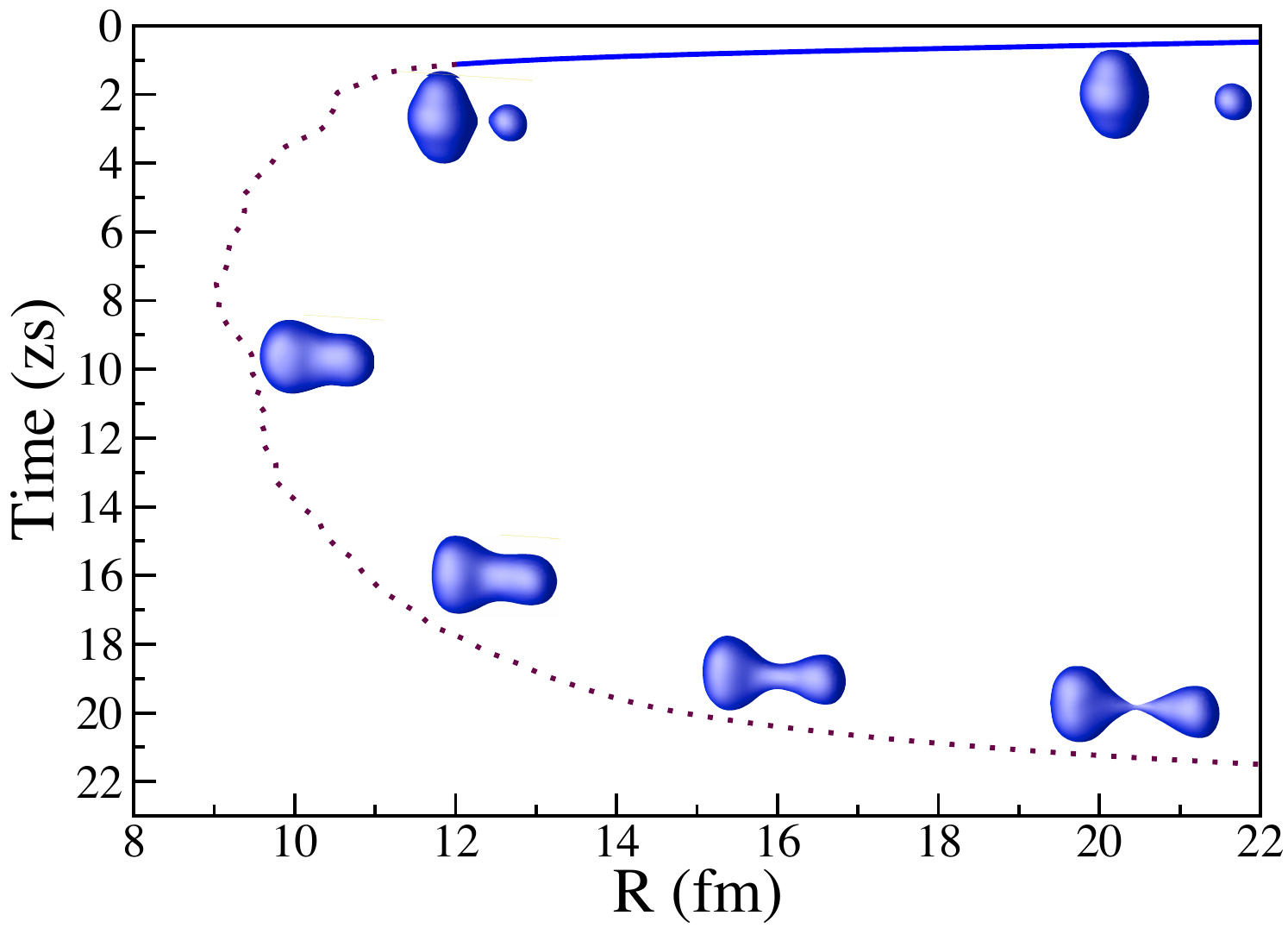}
\caption{Time evolution of the distance between the centers of mass of the fragments in the $^{40}$Ca+$^{238}$U central collision in the equatorial configuration at $E_{c.m.}=205.2$~MeV, i.e., $3\%$ above the frozen-HF barrier for this configuration.  The dotted line is indicative of the trajectory (see text).
The snapshots show densities at half the saturation density at different times.} 
\label{fig:Ca+U}   
\end{figure}

Figure \ref{fig:Ca+U} shows a trajectory of the fragments in a central collision of $^{40}$Ca+$^{238}$U at $E_{c.m.}=205.2$~MeV which is 
$3\%$ above the frozen-HF barrier for the equatorial configuration (i.e., with the   deformation axis of the $^{238}$U perpendicular to the collision axis). 
Note that the distance between the fragments is only well defined when they are well separated, hence the dotted line is only indicative. 
We see that 
the fragments re-separate after about 20~zs due to quasi-fission. 
Indeed, we also observe a large mass transfer from the heavy fragment to the lighter one, leading to  the average exit channel $^{103}$Mo+$^{175}$Yb \cite{sim12}. 
This partial equilibration of the mass asymmetry is a clear signature for quasi-fission. 

Systematic comparison of TDHF calculations with experimental data on quasi-fission are ongoing. 
Experimental signatures for quasi-fission can be found in the width of the fragment mass distribution \cite{she87,boc82}, in the total kinetic energy of the fragments \cite{itk11} and in their mass-angle distributions \cite{tok85,tho08}. 
The study of these quantities has enabled the characterisation of some important properties of the quasi-fission process, such as the role of the deformation \cite{hin96,nis12}, the effect of the magicity and of the $N/Z$ asymmetry in the entrance channel \cite{sim12b} as well as the influence of the shells in the exit channel \cite{koz10}. 
All these observations provide an ideal ground for testing theoretical models of nuclear dynamics including microscopic theories like TDHF on one hand, and macroscopic approaches \cite{zag06,kal11} on the other.

Finally, it is worth mentioning that very heavy systems such as actinide collisions could be considered an alternative way to form neutron rich heavy and super heavy nuclei \cite{zag06}. 
In actinide collisions, no fusion is expected and quasi-fission is by far the dominant process. 
However, TDHF calculations predict that some specific orientations of the nuclei may lead to an inverse quasi-fission process, where the mass transfer occurs from the lighter nucleus toward the heavier one \cite{gol09,ked10}. 
Note that shell effects in the $^{208}$Pb region may also favour the formation of a heavier complementary fragment \cite{zag06}. 

\subsection{Transfer reactions}

Let us now consider (multi-)nucleon transfer.
This field has attracted much experimental work in the recent past, in particular to infer the competition between sequential and cluster transfer with stable \cite{jia98,cor09,cor13,eve12,bis12} and exotic beams \cite{nav04,cha08,lem10,lem11}.

From a theoretical point of view, semi-microscopic approaches \cite{cor09} as well as purely microscopic approaches have been used. 
The TDHF theory treats the transfer mechanism at the independent particle level. 
Thus, it provides a microscopic benchmark for sequential transfer.
TDHF calculations of transfer reactions have then been performed with realistic interactions \cite{uma08,sim10,sek13}. 
In particular, by applying a particle number projection technique on the fragments \cite{sim10} transfer probabilities can be extracted \cite{sim10,sek13}. 

\begin{figure}[!tb]
\centering
\includegraphics[width=7cm]{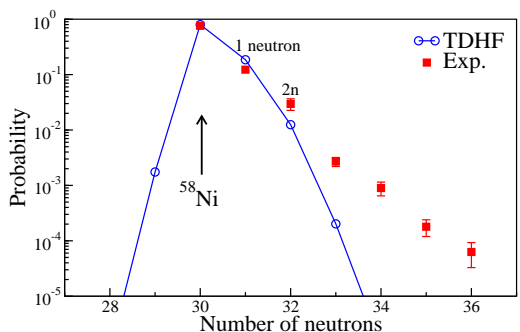}
\caption{Neutron number distribution of the light fragment in the exit channel of the $^{58}$Ni+$^{124}$Sn central collision at $E_{c.m.}=144.7$~MeV computed with TDHF (circles). 
The squares show experimental data extracted from Ref.~\cite{jia98} at $E_{c.m.}=153$~MeV$\simeq0.96V_B$ and $\theta_{c.m.}=127.5$~deg leading to the same distance of closest approach.
} 
\label{fig:proba}   
\end{figure}

Figure \ref{fig:proba} shows the neutron number distribution of the light fragment in the exit channel of the $^{58}$Ni+$^{124}$Sn central collision at $E_{c.m.}=144.7$~MeV computed with TDHF. 
The experimental data \cite{jia98} are from the same reaction at $E_{c.m.}=153$~MeV (i.e., about $4\%$ below the barrier $V_B\simeq160$~MeV \cite{jia98}) and $\theta_{c.m.}=127.5$~deg leading to the same distance of closest approach $R_0=13.93$~fm assuming Rutherford trajectories. 

We see that the order of magnitude for one and two neutron  transfer is well reproduced by the theory, although the 1n channel is overestimated while the 2n channel is underestimated. 
This can be attributed to an effect of pairing correlations, not included in the present calculations, favouring pair transfer. 
Scamps and Lacroix have recently included these correlations at the BCS level on top of a TDHF code \cite{sca13}. 
As a result, they obtained an enhancement of neutron pair transfer due to pairing correlations.

It is also clear from Fig.~\ref{fig:proba} that the transfer of more than two neutrons is underpredicted by one or several orders of magnitude. 
In fact, this is not surprising as the TDHF theory lacks beyond-mean-field fluctuations which need to be accounted for to reproduce the transfer of many particles. 
This is particularly crucial in deep-inelastic collisions. 
Such fluctuations have been included in the dynamics using a stochastic mean-field approach \cite{was09,yil11} or the time-dependent RPA \cite{sim11}. 
Indeed, it was shown that the fluctuations enhance the multi-nucleon transfer probabilities as compared to the independent particle picture. 

\section{Conclusions}

The time-dependent Hartree-Fock approach to many-body systems has been tested on nuclear dynamics. 
Both low-lying and high-lying collective vibrations are described. 
Only the direct decay of the latter is taken into account, which provides a good estimate of the width of the giant monopole resonance in light systems, but leads to an underestimation in heavy systems. 

The TDHF approach is also a powerful tool to investgate the complexity of the fusion process.
In particular, the coupling of the relative motion to all collective modes of the collision partners and of the compound system is automatically taken into account at the mean-field level. 

For reactions involving heavy nuclei, the fusion mechanism is in competition with quasi-fission of the di-nuclear system formed after capture. 
The quasi-fission process then prevents the formation of a compound nucleus.
TDHF calculations can be used to investigate the complex dynamics associated with quasi-fission.
In particular, the contact times between the fragments and the partial symmetrisation of their masses are characteristic of the quasi-fission process. 

Finally, the transfer of one or several nucleons has been discussed. 
Although the TDHF calculations reproduce the order of magnitude of the transfer probabilities of one and two nucleons, they underestimate the two-nucleon transfer channel and overestimate the one-nucleon transfer channel due to the lack of pairing correlations in the formalism. 
The transfer of many-nucleons requires the inclusions of beyond mean-field fluctuations, e.g., at the time-dependent RPA level. 

\section*{Acknowledgements}

B. Avez is thanked for providing his calculations with the \textsc{tdhfbrad} code. 
This work has been supported by the Australian Research Council by the Future Fellowship FT120100760, Laureate Fellowship FL110100098 and Discovery grants DP1094947 and DP110102858. 
The calculations have been performed on the NCI National Facility in Canberra, Australia, which is supported by the Australian Commonwealth Government.

\end{document}